\documentclass[iop]{emulateapj}

\usepackage{graphicx}
\usepackage{amsmath}

\shorttitle{Removing Large-scale Velocity}
\shortauthors{Cho}

\begin{document}

\title{ A technique of removing large-scale variations  in astronomical observations }
%\title{ Removing large-scale gradient and  obtaining small-scale fluctuations of observable quantities }

\author{ Jungyeon Cho}
\affil{Department of Astronomy and Space Science, Chungnam National University, Daejeon, Korea; jcho@cnu.ac.kr}
%\author{ Jungyeon Cho\altaffilmark{1},Hyunju Yoo\altaffilmark{2}}
%\affil{Department of Astronomy and Space Science, Chungnam National University, Daejeon, Korea}
%\altaffiltext{1}{jcho@cnu.ac.kr}
%\altaffiltext{2}{hyunju527@gmail.com}

\begin{abstract}
In many astrophysical systems, smoothly-varying large-scale variations coexist with
 small-scale  fluctuations.
For example, a large-scale velocity or density gradient can exist in molecular clouds
that exhibit small-scale turbulence.  
In redshifted 21cm observations, we also have two types of signals - the Galactic foreground emissions that change smoothly and the redshifted 21cm signals that change fast 
in frequency space.
Sometimes the  large-scale variations make it difficult to extract information on
small-scale fluctuations.
We propose a simple technique to remove smoothly varying large-scale variations.
Our technique relies on multi-point structure functions and can obtain
the magnitudes of small-scale fluctuations. It can also help us to filter out large-scale  
variations and retrieve small-scale maps.
We discuss applications of our technique to astrophysical observations.

\end{abstract}
\maketitle

\section{Introduction}
In many astrophysical fluids, smooth large-scale variations are overlaid with
fast-varying small-scale fluctuations.
For example, magnetic fields in molecular clouds may consist of
smoothly-varying mean components and shorter-scale turbulence components (see, for example, Girart et al. 2006; Hildebrand et al. 2009; Houde et al. 2009).
Velocity fields in molecular clouds also exhibit large-scale gradients, as well as small-scale
turbulent fluctuations (Imara \& Blitz 2011).
The separation of signals into large-scale and small-scale ones is not limited to spatial
fluctuations. In observations of the redshifted 21 cm lines, we may separate 
smoothly varying large-scale foreground components and fast-fluctuating small-scale cosmological components in frequency
space (Morales et al. 2006; Cho et al. 2012). We may also separate time-series data into two components.

If large-scale variations and small-scale fluctuations coexist, 
the large-scale components  sometimes make it difficult to obtain information on
small-scale fluctuations. 
For many applications, it is necessary to accurately measure small-scale fluctuations.
For example, we could constrain turbulence parameters by observing 
 the standard deviation, skewness, or kurtosis of 
column density (e.g., Burkhart et al. 2009).
If there is no large-scale variations of column density, it may be straightforward to
obtain those quantities from observations.
However, when there are large-scale variations, they will certainly affect all those quantities.

The situation is similar for centroid velocity (for an optically thin line), which is equal to the intensity-weighted average velocity (see Section \ref{sect:method}
for mathematical definition).
Centroid velocity contains information on turbulence velocity field and therefore has been used
to diagnose properties of interstellar turbulence (von Hoerner 1951; Dickman \& Kleiner 1985; Kleiner \& Dickman 1985; O'Dell
\& Casta\~{n}eda 1987; Miesch \& Bally 1994; Esquivel et al. 2007).
Therefore accurate measurements of the small-scale centroid velocity fluctuations is important for the study of interstellar turbulence.

Centroid velocity is also important
for measurement of interstellar magnetic fields
(Cho \& Yoo 2016; Gonz{\'a}lez-Casanova \& Lazarian 2017).
The Chandrasekhar-Fermi method (Chandrasekhar \& Fermi 1953) is a popular and simple
technique to obtain strengths of interstellar magnetic fields
projected on the plane of the
sky, which makes use of
polarized emission in FIR/sub-mm wavelengths from magnetically aligned grains 
(Gonatas et al. 1990; Lai et al. 2001; Di Francesco et al. 2001;
Crutcher et al. 2004; Girart et al. 2006; Curran \& Chrysostomou 2007; 
Heyer et al 2008; Mao et al. 2008; Tang et al. 2009; Sugitani et al. 2011; Pattle et al. 2017).
The Chandrasekhar-Fermi method is based on the following assumption: If the mean magnetic field
is strong, wandering of magnetic field lines is small and, therefore, variation of
polarization angle is small\footnote{
 In fact, the method assumes $\tan \delta \phi \sim \delta b/B_0$ ($\approx M_A$), where
 $\delta \phi$ is the variation of polarization angle, $\delta b$ is the strength of fluctuating magnetic
 field, $B_0$ is the strength of the mean magnetic field projected on the plane of the sky, and
 $M_A$ is the Alfv\'en Mach number.
}.
%two assumptions - energy equipartition
%between fluctuating velocity and fluctuating magnetic field ($(\delta v) \sim (\delta b)/\sqrt{ 4 \pi \rho }$), and $\tan (\delta \phi) \sim (\delta b)/B_0$ ($\equiv M_A$).
%Here 
Cho \& Yoo (2016) showed that, if there are $N$ independent eddies along the line of sight (LOS), the variation of polarization angle ($\delta \phi$) is reduced by $\sim \sqrt{N}$ due to random averaging effect. If we use the Chandrasekhar-Fermi method, the reduction in $\delta \phi$ results in overestimation of magnetic field strength
by a factor of $\sqrt{N}$.
Cho \& Yoo (2016) suggested
that the standard deviation of 
centroid velocity divided by average line width can tell us about $\sqrt{N}$ (see Cho (2017) for a  heuristic explanation for this).
%Cho (2017) provided a heuristic explanation for that and discussed how
%the number of independent eddies along the LOS affect observations, including measurement
%of  polarization angle variation and magnetic field strength in molecular clouds.
Therefore accurate measurements of the small-scale centroid velocity fluctuations is important for the application of the Chandrasekhar-Fermi method.
Since large-scale variations in the LOS velocity can severely
affect the standard deviation of centroid velocity, it is necessary to remove the large-scale 
LOS velocity variations.

%\bibitem[Burkhart et al.(2009)]{2009ApJ...693..250B} Burkhart, B., Falceta-Gon{\c c}alves, D., Kowal, G., \& Lazarian, A.\ 2009, \apj, 693, 250 

Fitting is frequently used to remove large-scale variations.
For example, magnetic fields in molecular cores frequently show an hour-glass morphology
(Schleuning 1998; Houde et al. 2004).
As explained in the previous paragraph,
the Chandrasekhar-Fermi method requires measurement of
$\delta \phi$.
However the large-scale magnetic morphology impedes
accurate measurement of the quantity, which makes it difficult to
apply the Chandrasekhar-Fermi method.
To model the hour-glass shape large-scale magnetic fields,
a fitting function of the form $x=g+gCy^2$, where $g$ and $C$ are constants, has been successfully used (Girart et al. 2006; Sugitani et al. 2010).
However, in many cases, fitting requires knowledge on the large-scale variations \textit{a priori}.

In this paper, we propose a technique to remove large-scale variations.
Our main goal is to obtain the standard deviations of small-scale quantities.
Nevertheless, our technique can be also used to filter out large-scale variations and retrieve small-scale maps.
Our technique requires multi-point structure functions and does not rely on fitting method.
We first describe theoretical backgrounds of our technique and 
numerical methods for testing our technique in Section 2.
We present our results in Section 3.
We give discussions and summary in Sections 4.

%consists of a large scale, slowly varying component with overlaid shorter scale fluctuations.

\section{Theoretical Considerations and numerical methods}

\subsection{Removal of large-scale variation with multi-point structure functions}
\begin{figure}
\center
\includegraphics[width=0.48\textwidth]{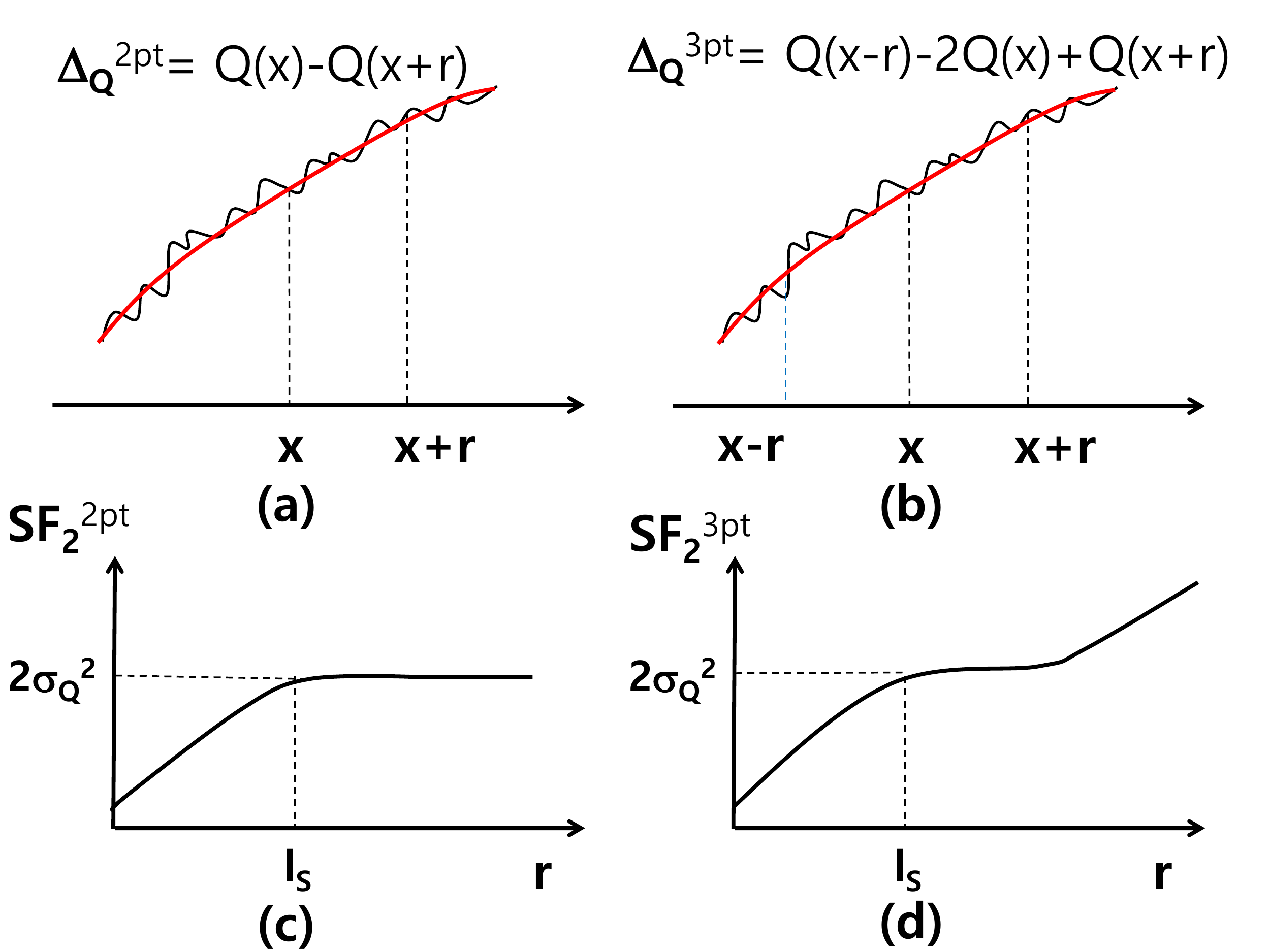}  
\caption{Large-scale variation and the 2-point and the 3-point structure functions.
(a) The large-scale variation dominates the quantity $\Delta_Q^{2pt}$ if the separation $r$ is large enough.
(b) It is possible that the small-scale fluctuations can dominate the quantity $\Delta_Q^{3pt}$ even if $r$ is large.
%$r> l_S$.
(c) The behavior of a second-order structure function in the absence of
     a large-scale variation.
 (d) The behavior of a \textit{multi-point} second-order structure function in the presence of
     large-scale variations. If the structure function successfully removes the large-scale effect,
     we will have an extended flat part (`plateau') on scales larger than $l_S$.
 }
\label{fig:1}
\end{figure}

Suppose that a quantity $Q$ in real space exhibits a large-scale variation, as well as 
small-scale fluctuations (see Figure \ref{fig:1}):
\begin{equation}
   Q({\bf x})=Q_L({\bf x})+Q_S({\bf x}).
\end{equation}
We assume the spatial average $\left< Q_S({\bf x}) \right>$ is zero when we calculate the average on scales larger than the small-scale correlation length $l_S$:
\begin{equation}
   \left< Q_S({\bf x}) \right> =0 \mbox{~~~~~~(if scale $ > l_S$)}.
\end{equation}
Our goal is to remove the large-scale variation $Q_L({\bf x})$ and obtain the standard deviation of the small-scale fluctuation
$\sigma_Q$.
In this subsection, we show that multi-point structure functions, rather
than the conventional 2-point structure function, can effectively remove the
large-scale variation.

%%%%%%%%%%%%%%%%%%%%%%%%%%
%%%%%%%%%%%%%%%%%%%%%%%%%%

\subsubsection{Two-point structure function}
In many problems, the usual 2-point second-order structure function for a variable $Q$,
\begin{equation}
  \mbox{SF}_2^{2pt}(r)=\left< | Q({\bf x}+{\bf r})-Q({\bf x})|^2 \right>_{\mbox{ avg. over \textbf{x}}},
\end{equation}
is frequently used to diagnose structures on different scales. In fact, it is related to power spectrum\footnote{
   If the one-dimensional power spectrum is proportional to
    $k^{-m}$ (i.e., $E(k) \propto k^{-m}$), where $k$
   is the wavenumber, the 2-point second-order structure function becomes
   SF$_2^{2pt}(r) \propto r^{m-1}$ (see, for example, Monin \& Iaglom 1975). 
   In the case of a steep spectrum (i.e., $m>3$), the 3-point structure function 
    (see Equation (\ref{eq:3ptsf}))
   should be used to reveal the correct spectral slope 
   (Falcon et al. 2007; Lazarian \& Pogosyan 2008; Cho \& Lazarian 2009). \label{foot:1}}.
However, we should be careful when we use SF$_2^{2pt}$.
In the presence of a large-scale variation, it may fail to reveal small-scale structures correctly.
%See, for example, Figure \ref{fig:1}(a).
If we select two points, \textbf{x} and \textbf{x}+\textbf{r}, as in Figure \ref{fig:1}(a),
the difference of the large-scale quantity
($|\Delta_L^{2pt}| \equiv |Q_L({\bf x})-Q_L({\bf x}+{\bf r}) |$) can be larger than
that of the small-scale quantity
($|\Delta_S^{2pt}| \equiv  |Q_S({\bf x})-Q_S({\bf x}+{\bf r})| $), which results in
\begin{multline}
 \mbox{SF}_2^{2pt} =\left< (\Delta_L^{2pt}+\Delta_S^{2pt})^2 \right>
             \approx \left< (\Delta_L^{2pt})^2 \right>
             \propto r^2  \\
             \mbox{~~~~(if $|\Delta_L^{2pt}|>|\Delta_S^{2pt}|$)},
\end{multline}
where $r\equiv |{\bf r}|$ and we assume that $\Delta_L^{2pt}$ varies smoothly.
Therefore, if the large-scale variation dominates the small-scale fluctuations, it is not possible to reveal statistics of the small-scale fluctuations from  SF$_2^{2pt}$.
%obtain $\sigma_Q$ from SF$_2^{2pt}$.

\subsubsection{Multi-point structure functions}
If we use multi-point structure functions, we can remove substantial amount of the large-scale effects.
Let us consider difference of $Q$ constructed with 3-points:
\begin{equation}
   \Delta_Q^{3pt}=Q({\bf x}-{\bf r})-2Q({\bf x})+Q({\bf x}+{\bf r})
\end{equation}
(see Figure \ref{fig:1}(b)).
It is trivial to show that $\Delta_Q^{3pt}$ can exactly eliminate a large-scale variation that has
a constant slope.
If the large-scale variation is so smooth that 
\begin{equation}
   |\Delta_L^{3pt}|<|\Delta_S^{3pt}|,
\end{equation}
as in Figure \ref{fig:1}(b), then the 3-point structure function  
\begin{equation}
  \mbox{SF}_2^{3pt} \equiv \frac{1}{3} 
                              \left< |Q({\bf x}-{\bf r})-2Q({\bf x})+Q({\bf x}+{\bf r})|^2 \right>,
                              \label{eq:3ptsf}
\end{equation}
can capture small-scale fluctuations correctly.

We can also construct 4-point and 5-point second-order structure functions as follows:
\begin{multline}
  \mbox{SF}_2^{4pt} \equiv \frac{1}{10} 
       \left< |Q({\bf x}-{\bf r})-3Q({\bf x})  \right. \\
      \left. +3Q({\bf x} +{\bf r}) -Q({\bf x}+2{\bf r})|^2 \right>, 
\end{multline}
\begin{multline}
  \mbox{SF}_2^{5pt} \equiv \frac{1}{35} 
       \left< |Q({\bf x}-2{\bf r})-4Q({\bf x}-{\bf r})+6Q({\bf x})  \right. \\ 
     \left.  -4Q({\bf x}+{\bf r})+Q({\bf x}+2{\bf r})|^2 \right>   
     \label{eq:5ptsf}
\end{multline}
(see Section \ref{sect:general} for general definition).
It is worth noting that SF$_2^{4pt}$ and SF$_2^{5pt}$ can exactly remove large-scale variations that follow a quadratic 
and a cubic polynomial, respectively.

\subsubsection{The behavior of structure functions}
Our primary goal is to obtain the standard deviation $\sigma_Q$ of small-scale fluctuations.
We may obtain $\sigma_Q^2$ from the shape of a second-order structure function.
Let us consider small-scale turbulence with correlation length $l_S$.
If there is \textit{no} large-scale variation, the behavior of a second-order structure function may look like
Figure \ref{fig:1}(c).
When the separation $r$ is small (i.e., when $r<l_S$),
the structure function reflects small-scale turbulence statistics and thus is
an increasing function of $r$ (see, Footnote \ref{foot:1}).
When 
$r > l_S$, $Q_S({\bf x}+{\bf r})$ and $Q_S({\bf x})$ are uncorrelated and
the second-order structure function gives
\begin{equation}
    \mbox{SF}_2 \approx 2\sigma_Q^2  \mbox{~~~~~~~ (if r $> l_S$)}.
    \label{eq:2sf}
\end{equation}
In fact, all the second-order structure functions mentioned above will give this value
when $r > l_S$.
 
On the other hand, if there is a large-scale variation and a multi-point structure function 
successfully removes a substantial part of it, then the behavior of the structure function will look like Figure \ref{fig:1}(d)\footnote{
       Note that, if the large-scale gradient is \textit{very small}, even the 2-point structure function 
       can also show a similar behavior.
}.
When  $r<l_S$, the structure function is an increasing function
of $r$.
When $r>l_S$ the structure function becomes flat, because the large-scale variation is
substantially removed.
The value of the second-order structure function for the flat part is $\sim 2\sigma_Q^2$.
As $r$ increases further, the accuracy of removing the large-scale variation by the structure function
gets worse and ultimately the large-scale variation makes the structure function increase again.
If the large-scale variation is poorly removed, the flat part will be very short.
The bottom line is that resolving the flat part (hereinafter `plateau') is essential for obtaining $\sigma_Q^2$.

\subsection{Numerical methods} \label{sect:method}
In the previous subsection, we argued that it is more advantageous to use multi-point SF$_2$'s 
to remove large-scale variations.
In this paper, we test this idea using numerical calculations.
The observable quantities we consider are the column density ($\Sigma$) 
and the velocity centroid ($V_c$).
The column density and the velocity centroid are defined by
\begin{eqnarray}
 \Sigma=\int \rho ~ dz, \label{eq:cden} \\
 V_c = \int \rho v_z ~dz \Bigg/ \int \rho ~dz,   \label{eq:v_c}
\end{eqnarray}
where $z$ is along the LOS, $\rho$ is the 3-dimensional (3D) density, and $v_z$ is the LOS velocity.
We consider two types of calculations.

\subsubsection{Simple sinusoidal large-scale variations}
We generate data that contain both small-scale fluctuations and a large-scale variation 
from the following procedure. We take turbulence data as small-scale fluctuations\footnote{
       Note that the small-scale data are not necessarily turbulence data. 
       We use existing small-scale turbulence data for simplicity.
}.

First, we generate 3D turbulence data from a direct numerical simulation
of isothermal supersonic magnetohydrodynamic (MHD) turbulence,
which contain only small-scale fluctuations.
The computational domain is a cubic box of size $2\pi$ ($\equiv L$) and 
consists of $512^3$ grid points.
The simulation is identical to the model `KF20' in Cho \& Yoo (2016).
 The driving scale is about 20 times smaller than the size of the computational domain,
which means that the typical size of largest energy-containing eddies is about 20 times smaller than the size of the computational box.
The sonic and the Alfv\'{e}nic Mach numbers are $\sim 7$ and $\sim 0.7$, respectively.
%The numerical resolution is $512^3$. 
The fluid velocity is zero ($v=0$),
density is one  ($\rho_0=1$), and the Alfv\'{e}n speed of mean field is one ($B_0/\sqrt{4 \pi \rho_0}=1$) at t=0.
Further description of the code can be found in Cho \& Yoo (2016).

\begin{figure}
\center
\includegraphics[width=0.48\textwidth]{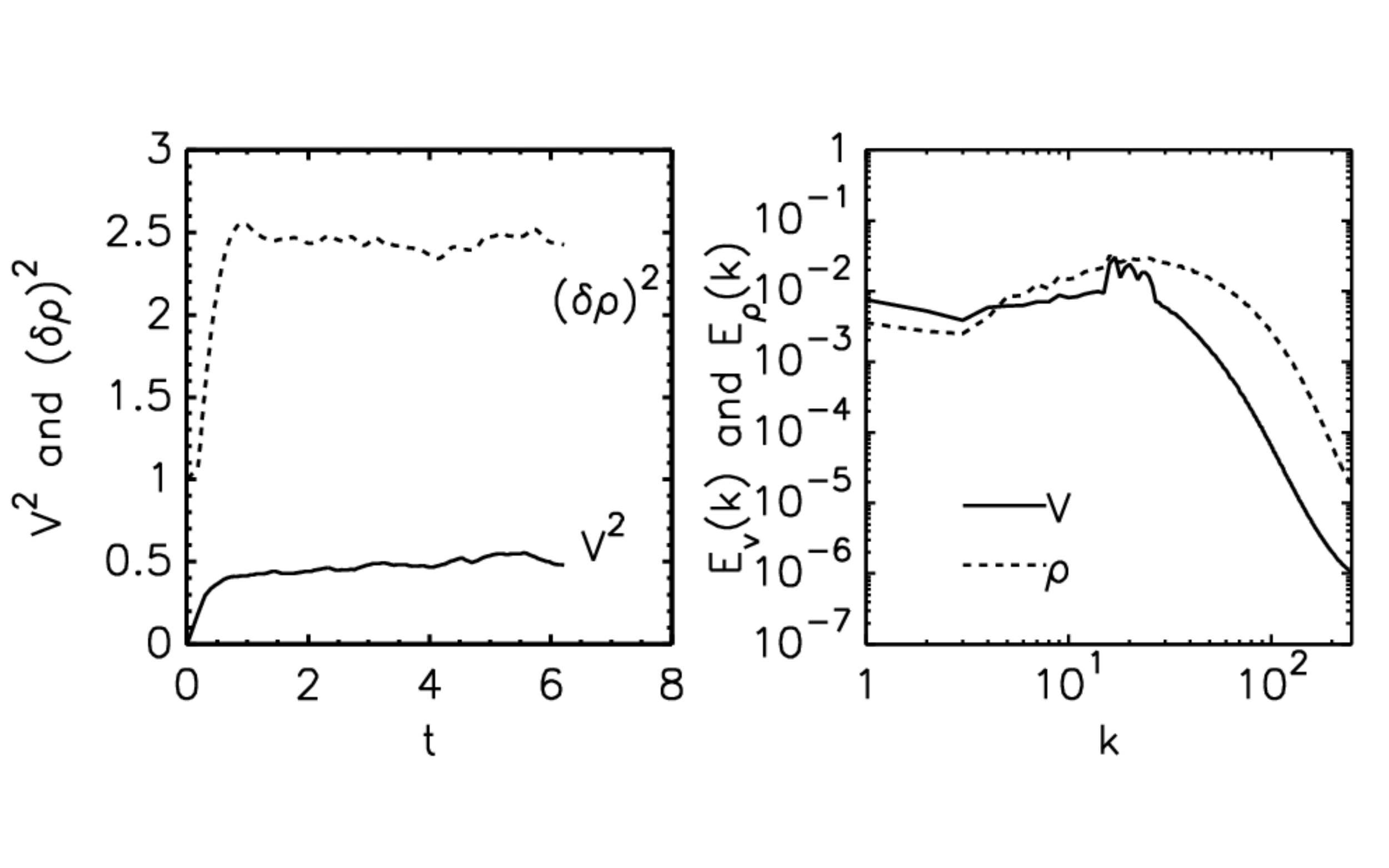}  
\caption{The Run K20. We drive a fluid at $k\sim 20$ and generate supersonic isothermal turbulence. We take data cubes from the run as small-scale fluctuating quantities.
(a) Time evolution of 3-dimensional (3D) $v^2$ and $(\delta \rho)^2$.
(b) The spectra of the 3D velocity ($v$) and density ($\rho$) at $t\sim 6$.
 }
\label{fig:2}
\end{figure}

Our goal is to obtain the magnitudes of small-scale fluctuations of
column density and velocity centroid.
%Here we focus on fluctuations in column density and velocity centroid (see Equations (\ref{eq:cden})
%and (\ref{eq:v_c})).
Since these fluctuations are related to 3D density and velocity
(see Equations (\ref{eq:cden})
and (\ref{eq:v_c})),
we plot time evolution of $v^2$ and $(\delta \rho)^2$
  in the left panel of Figure \ref{fig:2}, where
$v$ is the 3D velocity and $\delta \rho$ ($\equiv \rho-\rho_0$) is the fluctuating 3D density. 
The data we use are taken at t$\sim$6, at which the r.m.s. velocity is $\sim 0.7$ and  % t~6.2
$\delta \rho \sim 1.6$.
The right panel of Figure \ref{fig:2} shows spectra of the 3D velocity ($E_v(k)$) and
density ($E_\rho(k)$). They have peaks at $k\sim20$, which corresponds to
the average driving wavenumber.

Second, using the data, we calculate 
column density (Equation (\ref{eq:cden})) and centroid velocity (Equation (\ref{eq:v_c})).
The LOS is along the $z$-direction and is perpendicular to the mean magnetic field.
The standard deviations of column density and centroid velocity (\textit{without} a large-scale variation) along the LOS
are
\begin{equation}
     \sigma_\Sigma \approx 90,   \mbox{~~~~and~~~} \sigma_{V_c} \approx 0.084,
\end{equation}
and, therefore, we have
\begin{equation}
    2 (\sigma_\Sigma)^2 \approx 1.6\times 10^4,  \mbox{~~~~and~~~} 2(\sigma_{V_c})^2 \approx 0.014
    \label{eq:13}
\end{equation}
(see Table 1).
 
% $\delta\Sigma$ and $\delta V_c$

%%%%%%%%%%%%%%%%%%%%%%%%%%
%%%%%%%%%%%%%%%%%%%%%%%%%%

Third, after calculating column density and centroid velocity, we add simple large-scale variations. 
The large-scale variations have the sinusoidal form
\begin{equation}
       Q({\bf x}) = A_Q ~\sin\left[ k(x-\pi)\right],   \mbox{~~~k=1/2, 5/2, 9/2}, \label{eq:lsg}
\end{equation}
where %$x$ is along the mean magnetic field, 
$0<x\leq 2\pi$ and $Q$ is either 
%a line-of-sight velocity (for $V_c$) or 3D density (for column density $Sigma$).
 column density or centroid velocity.
The corresponding wavelengths of the large-scale variations are
       $\lambda$ ($=2\pi/k$) = 2L, 2L/5, and 2L/9, respectively.
The amplitude $A$ is 1024 for column density (i.e., $A_{\Sigma}$=1024) and 1.0 for centroid velocity (i.e., $A_{Vc}$=1.0), which are $\sim 10$ times larger than the amplitudes of the corresponding small-scale fluctuations.
We list properties of the turbulence data, including standard deviations of small-scale fluctuations ($\sigma_\Sigma$ and $\sigma_{V_c}$), in Table 1 (see Model KF20).

\subsubsection{More complicated turbulent large-scale variations}  \label{sect:ls}
In the previous subsection, we considered idealistic large-scale variations.
To see if our technique works also for more complicated large-scale fluctuations, we take 
large-scale turbulence data as the large-scale variations.
To be specific, we use data of isothermal turbulence driven at two different spatial scales simultaneously. The driving wavenumbers are near k$\sim$ 2.5 and k$\sim$20.
Since the two driving scales are well separated, we can assume that
the large-scale driving (i.e., driving near k$\sim$2.5) generates large-scale variations, while
the small-scale driving (i.e., driving near k$\sim$20) creates small-scale fluctuations.
We want to remove the former and retain the latter.
%In this case, fluctuations generated by the large-scale driving acts as large-scale variations, 
%while 
%those by the small-scale driving are the ones we want to retrieve.
The sonic Mach number is around unity and the numerical resolution is $512^3$.
The numerical setups for the simulation are virtually identical to those of
the Run CS\_L1.0\_S2.0 in Yoo \& Cho (2014), but the numerical resolution for the current run is higher.
We list properties of turbulence in Table 1 (see Model L1.0\_S2.0).

Since turbulence is driven at small and large scales simultaneously, it is not easy
to define which are small-scale fluctuations and which are large-scale ones.
Nevertheless, since our goal is to retrieve small-scale fluctuations, it is necessary
to have rough estimates about the magnitudes of small-scale fluctuations.
We calculate the standard deviations of the small-scale fluctuations, $\sigma_\Sigma $ and $\sigma_{V_c}$, from the following procedure.
First, we perform Fourier transformation of the real-space data and obtain wavevector-space data.
Second, we filter out large-scale data. To be specific, we set the Fourier amplitudes to zero when $k <10$ and retain the data when $k\geq 10$.
We take $k=10$ because the 3D spectra of velocity and density show different behaviors
for $k<10$ and $k>10$
(see Section \ref{sect:3.2} for details).
Third, we transform the filtered data back to real space.
Fourth, we calculate $\sigma_\Sigma $ and $\sigma_{V_c}$ from the (filtered) real-space data.
The resulting  $\sigma_\Sigma $ and $\sigma_{V_c}$ are
\begin{equation}
     \sigma_\Sigma \approx 16,   \mbox{~~~~and~~~} \sigma_{V_c} \approx 0.041,
\end{equation}
which give
\begin{equation}
    2 (\sigma_\Sigma)^2 \approx 510,   \mbox{~~~~and~~~} 2(\sigma_{V_c})^2 \approx 0.033
    \label{eq:2fk}
\end{equation}
(see the data for KF2.5\_20 in Table 1).

\section{Results}
\subsection{Sinusoidal large-scale variations}

%%%%%%%%%%%%%%%%%%%%%%%%%%
\begin{figure*}
\center
\includegraphics[width=0.70\textwidth]{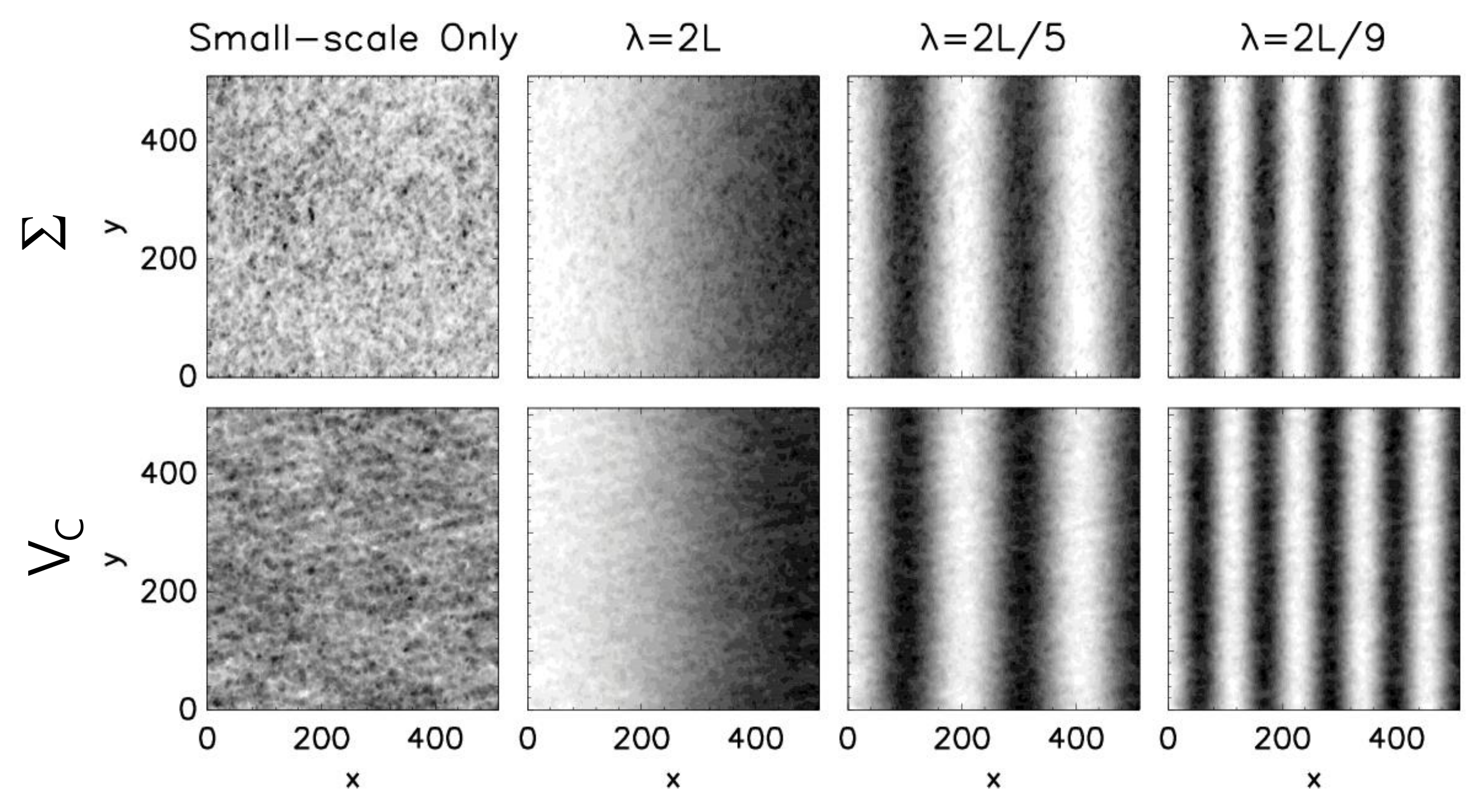}  
\caption{Contour plots of column density (upper panels) and centroid velocity (lower panels).
The far left panels (i.e. panels in the first column from the left) show only small-scale fluctuations.
Panels in the other columns contain both small-scale fluctuations, which are identical to
the ones in the first column, and large-scale variations of sinusoidal forms (see Equation (\ref{eq:lsg})). The large-scale variations dominate small-scale fluctuations. Note that $\lambda$ denotes wavelength of the large-scale variations and $L=2 \pi$.
 }
\label{fig:conto}
\end{figure*}
%%%%%%%%%%%%%%%%%%%%%%%%%%
%%%%%%%%%%%%%%%%%%%%%%%%%%
\begin{figure*}
\center
\includegraphics[width=0.70\textwidth]{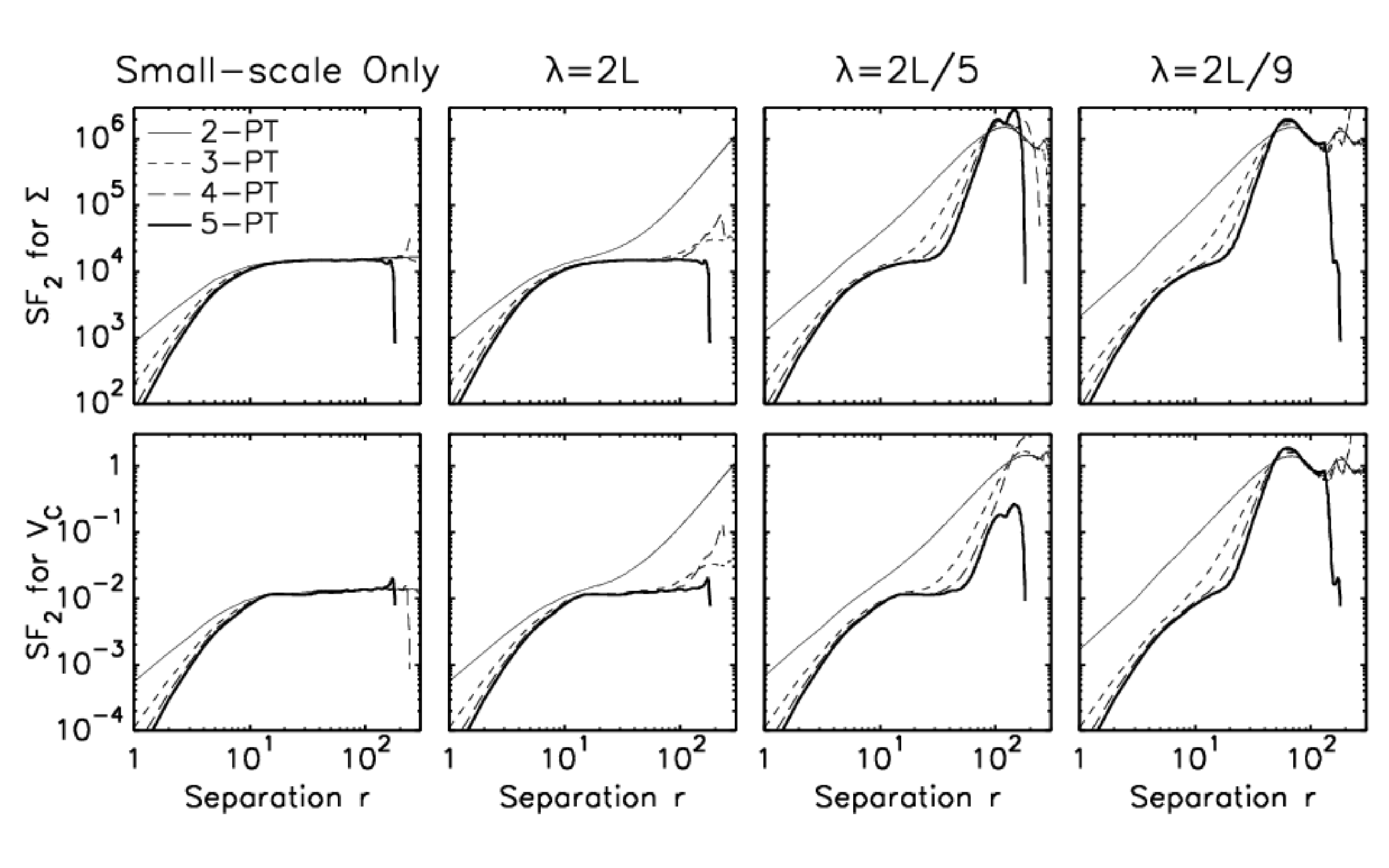}  
\caption{The multi-point second-order structure functions for
 column density (upper panels) and centroid velocity (lower panels).
 The arrangement of panels is the same as that of Figure \ref{fig:conto}.
 In the absence of large-scale variations (see the far left panels),
 all structure functions converge to a constant value for $r\gtrsim 10$.
 In the case of $\lambda=2L$ (panels in the second column from the left), the two-point structure functions (thin solid lines) monotonically increase, while other multi-point structure functions have  wide plateaus.
  In the case of $\lambda=2L/5$ (panels in the third column from the left), the 3-point structure functions (dashed lines) also monotonically increase, which means that they cannot remove
  the large-scale variations well.
  In the case of $\lambda=2L/9$ (far right panels), even the 5-point structure functions (thick solid lines) fail
  to resolve well-defined plateaus.
 Note that the 5-point structure function performs better than the other ones shown in the panels.
 }
\label{fig:k20}
\end{figure*}
%%%%%%%%%%%%%%%%%%%%%%%%%%

Figure \ref{fig:conto} shows maps for column density (upper panels) and centroid velocity (lower panels).
The far left panels (i.e., upper-left and lower-left panels) display maps
without a large-scale variation.
Since there is no large-scale variation, both column density and centroid velocity show
only small-scale fluctuations.
The panels in second, third, and last columns from the left display maps in the presence of large-scale variations with 
$\lambda=2L, 2L/5$, and $2L/9$, respectively.
As we can see in the maps, the large-scale variations of both column density and centroid velocity dominate small-scale ones.

Figure \ref{fig:k20} shows our main results - the multi-point second-order structure
functions.
The order of the panels is the same as that of Figure \ref{fig:conto}.
In case of small-scale fluctuations only (far left panels), 
all the structure functions are increasing functions of $r$ when $r\lesssim 10$ and
gradually approach the same constant value when $r > 10$,
which
is consistent with our expectation (see Figure \ref{fig:1}(c)).
The values of the structure functions for $r>10$ are 
\begin{equation}
       SF_2(r>10) \approx 1.5 \times 10^4   \mbox{~~~~~~(for  $\Sigma$)},
        \label{eq:16}
\end{equation}
 and
\begin{equation}
    SF_2(r>10) \sim 0.013  \mbox{~~~~~~ (for $V_c$)},  \label{eq:17}
\end{equation}
 which are
virtually identical to $2(\sigma_\Sigma)^2$ and $2(\sigma_{V_c})^2$, respectively (see Equation (\ref{eq:13}) and also Table 1).

In the presence of a large-scale variation with $\lambda$=2L 
(i.e., $k=1/2$; see Equation (\ref{eq:lsg})), all structure functions, except
SF$_2^{2pt}$ (thin solid curves),
 can resolve the flat part (`plateau') quite well (see the panels in the second column from the left). 
 %Note however that $SF_2^{2pt}$ (thin solid curves) does not resolve the
 %plateau well.
The panels in the right two columns show that, 
when the wavelength $\lambda$ of the large-scale variation becomes smaller, 
SF$_2^{2pt}$ can no longer resolve the plateau.
When $\lambda=2L/5$ (the third panels from the left),  SF$_2^{4pt}$ and
SF$_2^{5pt}$ clearly resolve the plateau, while $SF_2^{3pt}$ can barely resolve it.
When $\lambda=2L/9$ (far right panels), the multi-point structure functions marginally resolve the flat part.
Among the multi-point structure functions shown in the panels, SF$_2^{5pt}$ performs best.
The values of the second-order structure functions at the plateau are very
close to the values in Equation (\ref{eq:13}), or Equations (\ref{eq:16}) and (\ref{eq:17}), which
means that we can indeed extract $\sigma_\Sigma^2$ or $\sigma_{V_c}^2$ using the multi-point
structure functions.

\subsection{Turbulent large-scale fluctuations} \label{sect:3.2}

%%%%%%%%%%%%%%%%%%%%%%%%%%
\begin{figure*}
\center
     \includegraphics[width=0.90\textwidth]{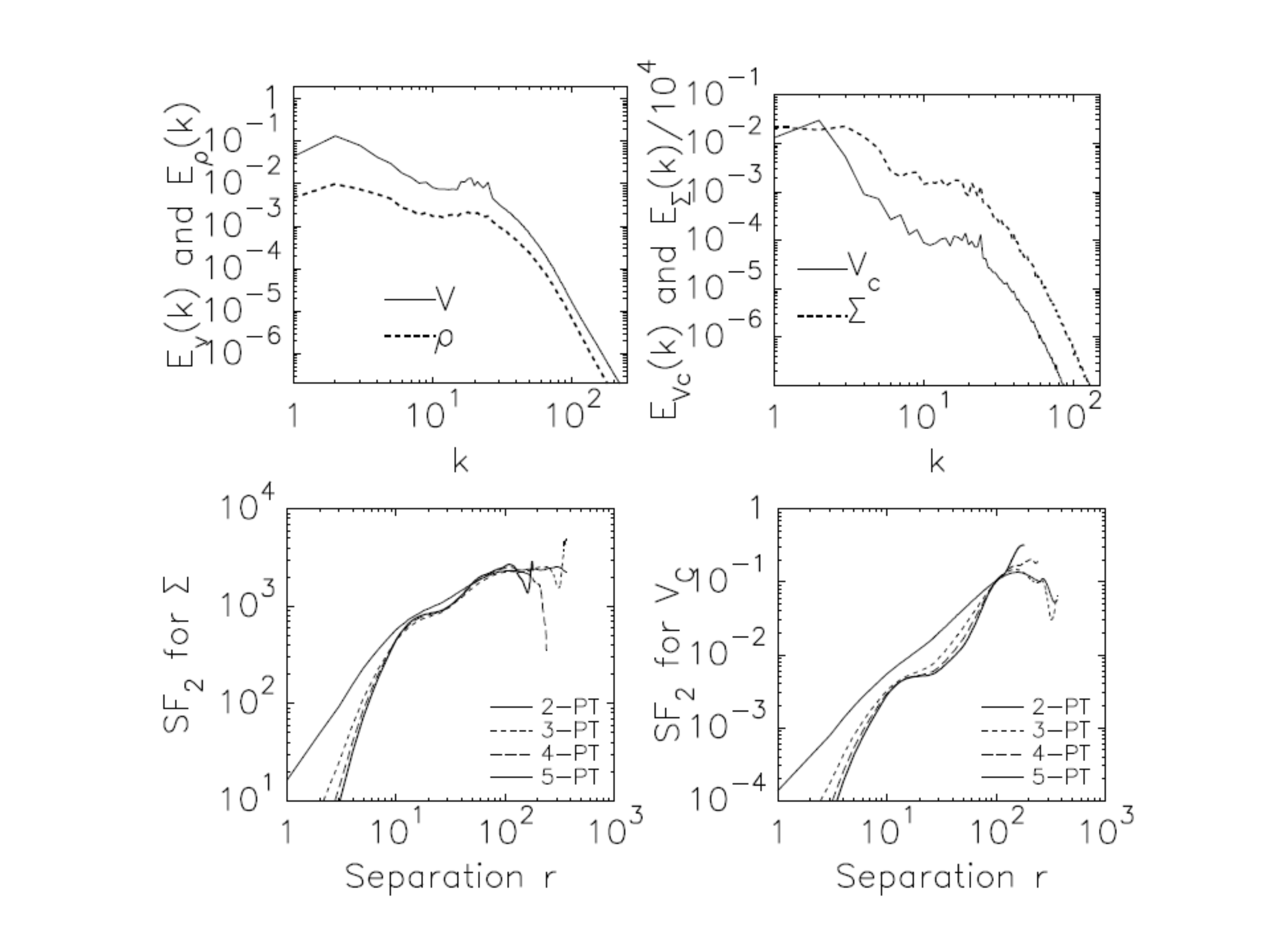}   
\caption{The Run K2.5\_20.
We drive the fluid at $k\sim 2.5$ and $k\sim 20$ simultaneously and generate transonic isothermal turbulence. We regard the structures generated by the large-scale driving (i.e., $k\sim 2.5$)
as large-scale variations and the ones by the small-scale driving (i.e., $k\sim 20$) 
as small-scale fluctuations.
\textit{Upper-left:} Spectra of (3D) $v$ and $\rho$.
\textit{Upper-right:} Spectra of (2D) column density $\Sigma$ and centroid velocity $V_c$.
\textit{Lower-left:} Second-order structure functions for $\Sigma$.
\textit{Lower-right:} Second-order structure functions for $V_c$.
Note the plateaus near $k\sim 15$.
 }
\label{fig:ls}  % large-small
\end{figure*}
%%%%%%%%%%%%%%%%%%%%%%%%%%

As explained in Section \ref{sect:ls}, we apply
the multi-point structure functions to 
data of isothermal turbulence driven simultaneously at two different spatial scales.
We plot the results in Figure \ref{fig:ls}: spectra of the 3D velocity and density (upper-left panel),  spectra of the 2D column density and centroid velocity (upper-right panel),
structure functions of column density (lower-left panel), 
and those of centroid velocity (lower-right panel).

The spectra of 3D velocity and density (upper-left panel) clearly show two peaks, one
near the average wavenumber of large-scale driving ($k\sim 2.5$) and the other
near the average wavenumber of small-scale driving ($k\sim 20$).
The large-scale fluctuations of 3D velocity and density 
exhibit roughly power-law spectra for $k<10$.
Both spectra get flatter after $k\sim10$ and the effects of the small-scale driving
become clearly visible for $k \gtrsim 15$.
The behavior of the spectra of column density and centroid velocity (upper-right panel) is also
similar. They decrease as the wavenumber increases for $k<10$,
become flat for $10 \lesssim k \lesssim 20$, and decrease again after $k\sim 20$.
We may assume that the flat and decreasing spectra for $k\gtrsim 10$ are due to
small-scale fluctuations.

The 2-point structure functions ($SF_2^{2pt}$) in the lower panels do not exhibit plateaus, while
structure functions based on 3 or more points clearly show plateaus.
The values of the multi-point second-order structure functions at the plateaus are 
\begin{equation}
     SF_2(\mbox{at plateau})\sim 800 \mbox{~~~~~~(for $\Sigma$)}
\end{equation}
and
\begin{equation}
     SF_2(\mbox{at plateau})\sim 0.005 \mbox{~~~~~~ (for $V_c$)},
\end{equation}
which are not far from the estimates for $2(\sigma\Sigma)^2$ and $2(\sigma_{V_c})^2$, respectively, in Equation (\ref{eq:2fk}).
Therefore we can conclude that the multi-point structure functions can also
remove complicated large-scale variations reasonably well.

\section{Discussions and Summary}

%%%%%%%%%%%%%%%%%%%%%%%%%%
\begin{figure*}
\center
\includegraphics[width=0.70\textwidth]{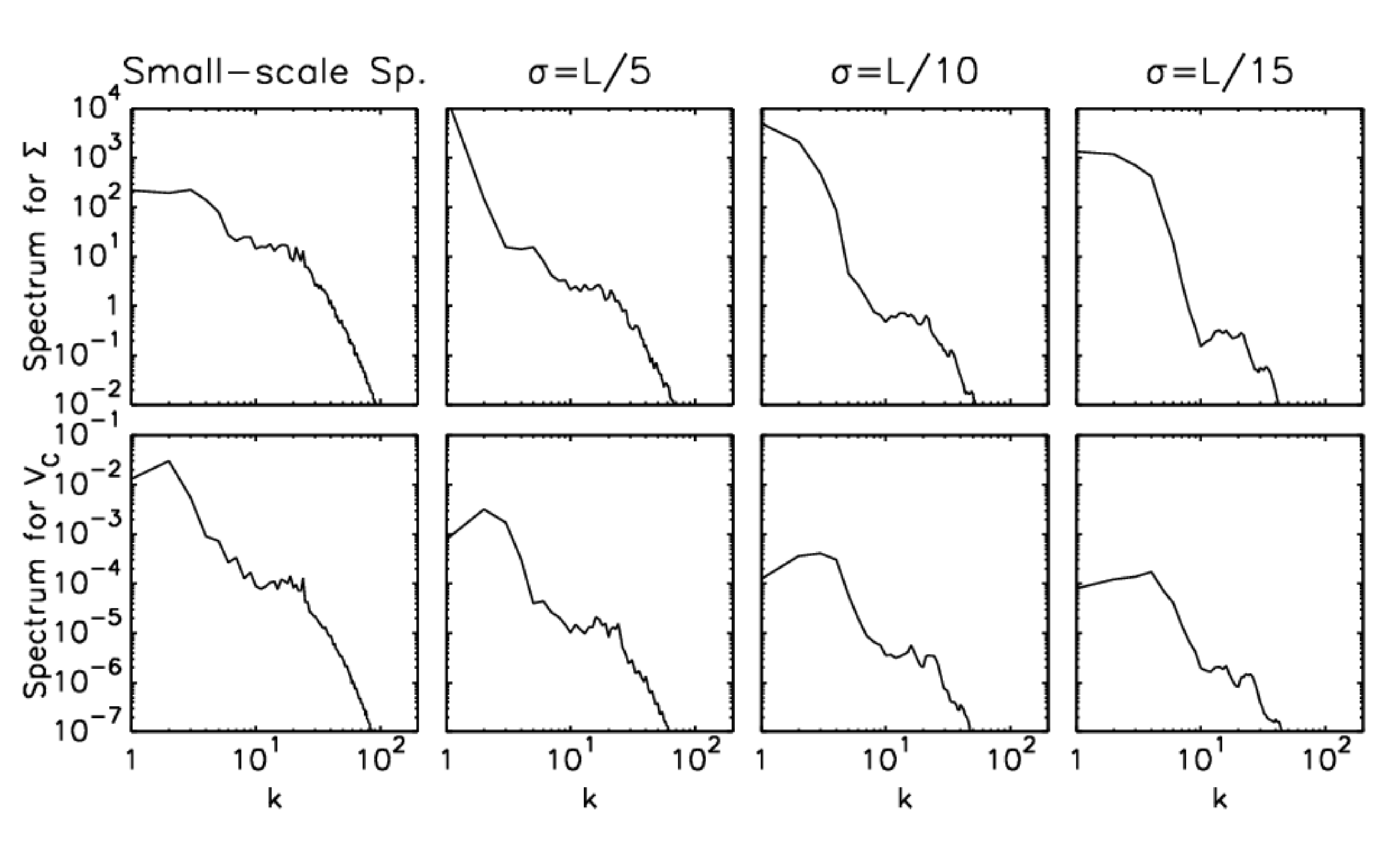}  
\caption{Spectra for column density (upper panels) and centroid velocity (lower panels).
We use a data cube from the Run K2.5\_20.
We calculate spectra using either the original maps (512 $\times$ 512) 
or partial maps (256 $\times$ 256) .
The reason we use the partial maps is to include the edge effect.
 The plots in the first column from the left are the spectra of the original maps
 (on a grid of 512 $\times$ 512).
 The plots in the other columns are the spectra of the partial maps 
 (on a grid of 256 $\times$ 256) tapered by gaussian windows with different widths (see 
 the standard deviations $\sigma$'s of the window functions
 on the panels).
 }
\label{fig:sp}
\end{figure*}
%%%%%%%%%%%%%%%%%%%%%%%%%%

\subsection{Spectrum vs. multi-point structure functions}
Power spectrum is also a useful tool to study small-scale fluctuations.
Indeed, if we can obtain the correct power spectrum, it may be possible to
separate large-scale variations and small-scale fluctuations.
%For example. it is not difficult to estimate the spectra of
%large-scale and small-scale fluctuations from the shapes of power spectra in the upper-left panel of Figure \ref{fig:ls}.
However, obtaining the correct spectrum is not easy when the data are not periodic.
If the data are not periodic, the discontinuity at the edge can severely affect the shape of
the power spectrum.
To reduce this artifact, a tapering window function is frequently used, which forces
the values near the edge converge to zero.
While the tapering method should work fine when there are only small-scale fluctuations,
it may cause nontrivial effects when there are also large-scale variations.

To demonstrate the effects of tapering window, 
we calculate power spectra of non-periodic 2D maps
using gaussian tapering windows.
We make use of the column density and the centroid velocity maps of the Run K2.5\_20, the resolution of which is 512 $\times$ 512.
In order to make the maps non-periodic we divide each map into 4 equal quadrants and take only one of them,
the resolution of which is 256 $\times$ 256. 
To be precise, the original periodic maps are define for $0 < x,y \leq 2\pi$ and the new non-periodic maps
  are defined for  $0 < x,y \leq \pi$.
We apply 2D gaussian tapering windows with different widths
\begin{equation}
    W(x,y)=e^{ [(x-\pi/2)^2+(y-\pi/2)^2]/(2 \sigma^2)},  
\end{equation}
%where $0 < x,y \leq L/2$ ($\equiv \pi$), 
where $\sigma$=L/5, L/10, and L/15,
to the non-periodic maps and calculate spectra.
We plot the results in Figure \ref{fig:sp}.
The upper and lower panels are for column density and centroid velocity, respectively.
The far left panels show the spectra of the original maps (with 512 $\times$ 512 resolution),
which should be identical to the spectra in the upper-right panel of Figure \ref{fig:ls}. %(b).
Note that each spectra has two components - one for $k\lesssim 10$ and the other
for $k\gtrsim 10$.
The spectra in the other columns are the results of 2D gaussian tapering.
From left to right, the standard deviation ($\sigma$) of the gaussian function decreases.
%In the case of $\sigma$=L/5 (see the panels in the second column), the edge effect may be still important and, therefore, it is difficult to distinguish 
%two components in each spectrum. 
%In the cases of $\sigma$=L/10 and $L/15$, the small-scale component seems to be marginally visible.
In all the cases with the tapering windows, the small-scale component seems to be marginally visible.
Nevertheless it may be difficult to draw any useful  information from the spectra.

As we can see in Figure \ref{fig:sp}, the shape of spectrum changes when the shape of
the tapering window changes.
It may be possible to get a correct power spectrum if we know a proper shape of the window function.
However, there is no way to know the proper shape of the window function \textit{ a priori}.
The bottom line is that, although spectrum provides useful information on power distribution as 
a function of scale, it is not easy to obtain the correct spectrum.
On the other hand, the multi-point structure functions do not require any knowledge 
\textit{a priori}, which makes them 
 more useful
 in deriving information on small-scale fluctuations.

%%%%%%%%%%%%%%%%%%%%%%%%%%
\begin{figure*}
\center
\includegraphics[width=0.70\textwidth]{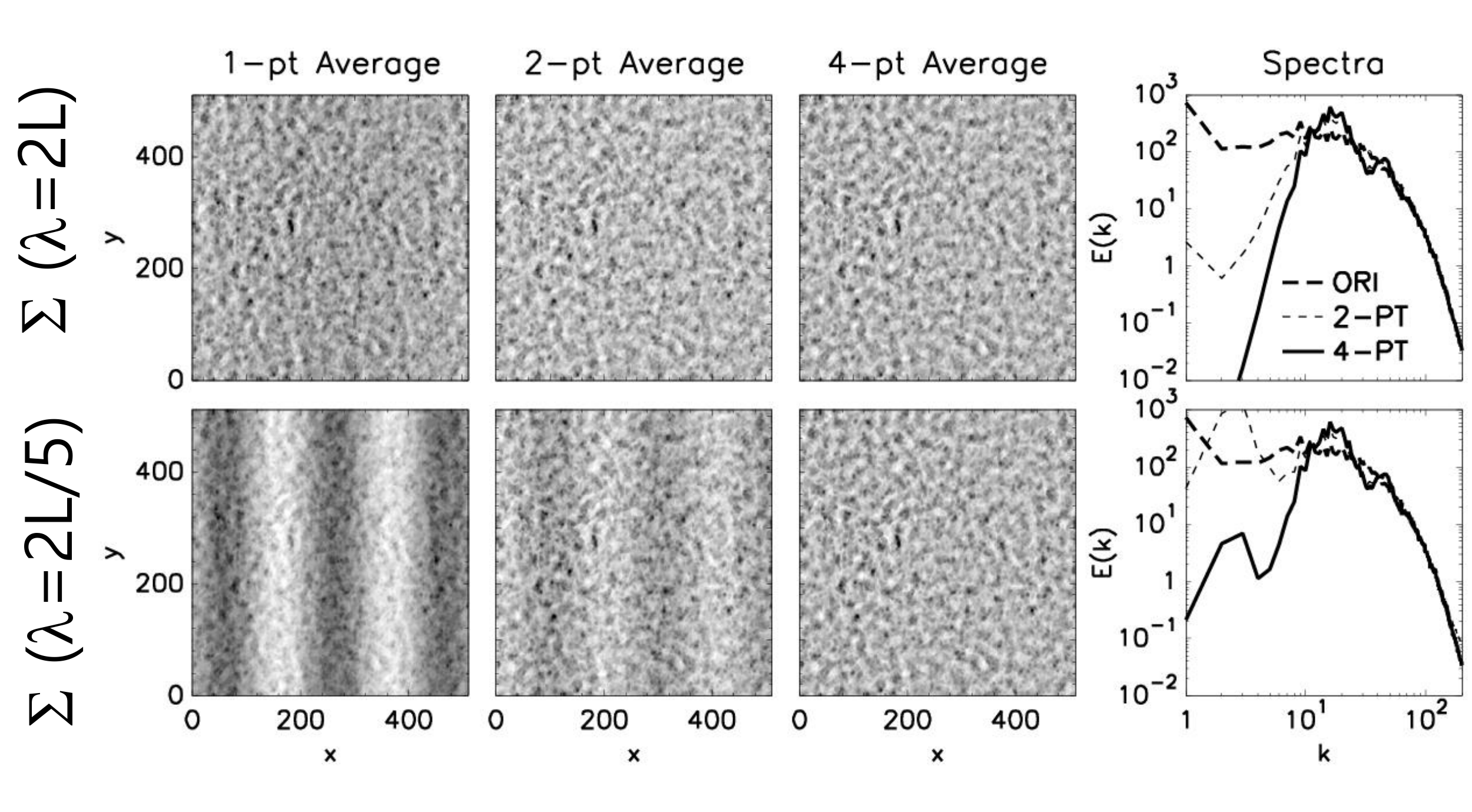}  
\caption{The reconstructed small-scale column density maps from the multi-point averaging technique
for $\lambda=2L$ (upper panels) and  $2L/5$ (lower panels).
We filter out the large-scale variations (see the maps in the second and third columns from the left in Figure \ref{fig:conto}) using the usual
(1-point) average (Equation (\ref{eq:usual})),
the 2-point average (Equations (\ref{eq:2ptavg})), and the 4-point average (Equation (\ref{eq:4ptavg})).
The contour plots in the first, second, and third column from the left are maps reconstructed with 
the usual (1-point) averaging method, 
 the 2-point averaging technique, and the 4-point averaging technique, respectively.
The plots in the far right column show spectra.
The thick long-dashed lines are for the original small-scale spectrum, which should be identical to the one in the upper-left panel of Figure \ref{fig:sp}.
The dashed and the solid lines denote the spectra of the maps reconstructed with the 2-point and the 4-point average techniques, respectively.
Note that all 3 spectra coincide well on small scales, which means that
the multi-point average technique can be used to reconstruct small-scale maps and spectra.
 }
\label{fig:map}
\end{figure*}
%%%%%%%%%%%%%%%%%%%%%%%%%%

%However, in case the data are not periodic, it has limited applicability. edge-effect.

\subsection{Obtaining a small-scale map}  \label{sect:smap}
Our technique discussed in earlier sections returns only the magnitudes of small-scale fluctuations.
In this subsection, we demonstrate our technique can be also used to filter out large-scale variations and obtain a small-scale map.
For simplicity, we use the 3-point (SF$_2^{3pt}$) 
and the 5-point (SF$_2^{5pt}$) second-order
structure functions.

Suppose that we have a map of an observable quantity $Q$ that contains both large-scale variations ($Q_L$) and small-scale fluctuations ($Q_S$).
If SF$_2^{3pt}$ or  SF$_2^{5pt}$
shows 
a plateau near a scale $r_p$, then we have
\begin{equation}
   Q_L({\bf x})\approx \left[ Q_L({\bf x}+{\bf r}) + Q_L({\bf x}-{\bf r}) \right]/2
\end{equation}
for SF$_2^{3pt}$ and
\begin{multline}
   Q_L({\bf x})\approx \left[ 4Q_L({\bf x}+{\bf r}) + 4Q_L({\bf x}-{\bf r}) \right. \\
      \left.  -Q_L({\bf x}+2{\bf r})-Q_L({\bf x}-2{\bf r})  \right]/6
\end{multline}
for SF$_2^{5pt}$
(see the definitions of SF$_2^{3pt}$ and SF$_2^{5pt}$), where ${\bf x}$ is a point on the map,
\textbf{r} is a 2D displacement vector, and $|{\bf r}|\sim r_p$.
Therefore, the 2-point average
\begin{equation}
 \bar{Q}({\bf x}) = \sum_{r_p-\Delta < |{\bf r}| < r_p+\Delta} \left[ Q_L({\bf x}+{\bf r}) + Q_L({\bf x}-{\bf r}) \right]/2N
  % \mbox{~~(for SF$_2^{3pt}$)}  
  \label{eq:2ptavg}
\end{equation}
and the 4-point average
\begin{multline}
  \bar{Q}({\bf x})= \sum_{r_p-\Delta < |{\bf r}| < r_p+\Delta} \left[ 4Q_L({\bf x}+{\bf r}) + 4Q_L({\bf x}-{\bf r}) \right.  \\
      \left.  -Q_L({\bf x}+2{\bf r})-Q_L({\bf x}-2{\bf r})  \right]/6N
      \label{eq:4ptavg}
\end{multline}
should be very good approximations for $Q_L({\bf x})$.
Here both $r_p-\Delta$ and $r_p+\Delta$ should lie in the plateau scale and
$N$ is the number of summation.
Note that the multi-point averages are different from the usual (1-point) average with a top-hat window:
\begin{equation}
   \bar{Q}({\bf x}) = \sum_{ |{\bf x}-{\bf x}^{\prime}|<r_p } Q_L({\bf x}^{\prime})/N.
   \label{eq:usual}
\end{equation}
If we calculate a multi-point overage on a scale smaller than the plateau scale,
the the value $\bar{Q}({\bf x})$ contains part of small-scale fluctuations.
On the other hand, if we calculate a multi-point overage on a scale larger than the plateau scale,
then the value $\bar{Q}({\bf x})$ loses some information about large-scale fluctuations. 

After obtaining an approximate value of  $Q_L({\bf x})$ (i.e., $\bar{Q}({\bf x})$),  it is trivial to obtain the
small-scale value $Q_S({\bf x})$:
\begin{equation}
   Q_S({\bf x}) \approx Q({\bf x})-\bar{Q}({\bf x}).
\end{equation}
We may calculate spectrum of small-scale fluctuations using $Q_S({\bf x})$.

In Figure \ref{fig:map} we demonstrate that this procedure is indeed working.
We apply the multi-point average technique to the column density maps shown in Figure \ref{fig:conto}, in which we can clearly see that the large-scale variations dominate the small-scale fluctuations.
We plot the results for the cases of $\lambda=2L$ and $\lambda=2L/5$ in Figure \ref{fig:map}.
We use $r_p=17.5$ and $\Delta=2.5$ (see Equations (\ref{eq:2ptavg}) and (\ref{eq:4ptavg})).
Note that, while both SF$_2^{3pt}$ and SF$_2^{5pt}$ for $\Sigma$ have wide plateaus 
for $\lambda=2L$, only SF$_2^{5pt}$ has a reasonably wide plateau near $r=17.5$ for
$\lambda=2L/5$ (see Figure \ref{fig:k20}).
We plot the resulting small-scale maps of the usual 1-point average (Equation (\ref{eq:usual})), the 2-point average (Equation (\ref{eq:2ptavg})), and the 4-point average (Equation (\ref{eq:4ptavg})) in the first, the second, and the third column from the left, respectively.
The upper panels are for $\lambda=2L$ and the lower panels are for
$\lambda=2L/5$. 
%We apply the multi-point averaging technique to the maps in the far-left column.
As we can see in the contour plots, since the large-scale variation is smooth enough
in the case of $\lambda=2L$ (upper panels), all 3 averaging methods can remove the large-scale
variation quite well. 
However, in the case of $\lambda=2L/5$ (lower panels), the usual 1-point average and the 2-point average
leave residuals of the large-scale variation on the maps, which means the usual 1-point average and the 2-point average cannot filter out the large-scale
variation completely. 
The result of the usual 1-point average is worse than that of the 2-point average.
On the other hand, filtering by the 4-point average does not leave visible residuals on the map
(see the lower panel in the third column from the left).
These results are not surprising because the 5-point structure function does have a well-defined
plateau near $r\sim 17.5$, while the 3-point structure function doesn't.
%\footnote{
%Our results that are not shown in this paper also support this statement.
%For $\lambda=2L$, both the 3-point and the 5-point structures have extended plateaus 
%(see Figure \ref{fig:k20}) and the 3-point and the 5-point average techniques produce small-scale maps without notable tracers of large-scale variations. 
%However, in the case of $\lambda=2L/9$, the 5-point structure function does not have a well-defined plateau and, hence, even the 5-point average technique leaves residuals of large-scale variations.
%}.

The line plots is in far right panels show the power spectra.
The thick long-dashed lines in the upper and the lower panels denote the spectrum of 
the original small-scale map of column density  
(see the upper-left panel of Figure \ref{fig:conto} for the original small-scale map).
The dashed and the thick solid lines represent the spectra of the small-scale maps obtained by the 2-point and the
4-point average techniques, respectively.
That is, they are spectra of the maps in the second and third columns in Figure \ref{fig:map}.
The spectra represented by the dashed and the thick solid lines do not
have significant powers at small wavenumbers (i.e., $k\lesssim 10$).
However, the spectrum represented by the dashed line in the lower panel clearly shows a peak near $k\sim 2.5$, which
corresponds to the wavenumber of the large-scale variation.
Note that the values of $E(k)$  is largest at $k=2.5$ for the dashed line, which is
in agreement with the fact that the residual of the large-scale variation is an outstanding
feature of the map in second-lower panel from the left.
The thick solid curve in the lower panel also has a peaks near $k\sim 2.5$.
But, its value  at $k=2.5$ is not large,
which is consistent with the fact that the residual of the large-scale variation is not really visible
on the map in the third-lower panel from the left.
It is worth noting that the spectra from the 2-point and the 4-point average techniques
virtually coincide with the spectrum of the original map when the wavenumber $k$ is large.

\subsection{Application to observations}
In this paper, we have proposed and tested a technique to remove 
large-scale variations and obtain magnitudes of small-scale fluctuations.
Our technique does not rely on fitting method that requires knowledge on
a fitting function \textit{a priori}.
Although we have focused only on column density and centroid velocity in this paper,
we can also apply our technique to FIR/sub-mm polarization, redshifted 21 cm observations, 
 or synchrotron emission data.
 In principle, our technique is applicable to any data that contain large-scale and small-scale fluctuations, if their spatial/temporal/frequency scales are well separated.
 For example, we can use our technique to separate small-scale fluctuating velocity and
 large-scale rotational velocity.
 We can also use our technique to obtain variations of
 polarization angles in regions where magnetic fields have hourglass morphologies.

\subsection{Construction of an n-point structure function} \label{sect:general}
In general, we can construct an n-point second-order structure function as follows:
\begin{equation}
     SF_2^{n-pt}(r)=\left< | \Delta^{n} |^2 \right>
     \label{eq:general}
\end{equation}
with
\begin{equation}
  \Delta^{n} = \frac{1}{\mathcal{N}}
           \sum_{l=0}^{n-1} (-1)^{l} { n-1 \choose l} Q\left(x+ (\frac{n-1}{2}-l)r \right)
\end{equation}
with
\begin{equation}
  \mathcal{N}=\frac{1}{2}\sum_{l=0}^{n-1} {n-1 \choose l}^2.
\end{equation}
Here ${n \choose l}$ is the binomial coefficient and $(n-1)/2$ can be either $n/2$ or $n/2-1$ if
$n$ is an even number.
Note that $\Delta^{n}$ is the same as the n-th order central difference.

\subsection{Summary}
In summary, we have obtained the following results.
\begin{enumerate}
  \item We develop a technique that can remove large-scale variations in observable quantities.
  Our technique relies on multi-point structure functions and gives us magnitudes of small-scale 
      fluctuations (see Equations (\ref{eq:3ptsf})-(\ref{eq:5ptsf}), and (\ref{eq:general})).
      
      \item Our technique works fine for a large-scale variation of a simple sinusoidal form. 
       It also works reasonably well for a more complicated turbulent large-scale fluctuations.
         
      \item If a second-order structure function shows a plateau, then the variance
            of the small-scale fluctuations is equal to the value of the structure function at
            the plateau divided by two (Equation (\ref{eq:2sf})).
   
   \item Our technique can be used to separate small-scale fluctuations and large-scale variations. 
           We have discussed how to filter out large-scale variations and obtain maps of
   small-scale fluctuations using multi-point averages (Section \ref{sect:smap}).

\end{enumerate}

\acknowledgements
This  work is supported by the National R \& D Program through 
the National Research Foundation of Korea Grants funded
by the Korean Government
 (NRF-2016R1A5A1013277 and NRF-2016R1D1A1B02015014).
We thank  Hyunju Yoo for providing the data cube of the Run KF2.5\_20.
We also thank Min-Young Lee for useful discussions.

%\newpage
\begin{deluxetable}{cccccccccccc}
\tabletypesize{\scriptsize}
%\rotate
\tablecaption{Simulations.}
\tablewidth{0pt}
\tablehead{
     \colhead{Run} 
   & \colhead{Resolution} 
   & \colhead{$M_s$ \tablenotemark{a}}
   & \colhead{$B_{0}/\sqrt{ 4 \pi \bar{\rho}}$ \tablenotemark{b}} 
   & \colhead{$k_f$ \tablenotemark{c}} 
   & \colhead{$\sigma_{V_c}$ \tablenotemark{d} }
   & \colhead{$2(\sigma_{V_c})^2$ }
   & \colhead{$\sigma_\Sigma$ \tablenotemark{e} }
   & \colhead{$2(\sigma_\Sigma)^2$ }
   & \colhead{A$_{\Sigma} $ \tablenotemark{f} }
   & \colhead{A$_{Vc} $ \tablenotemark{g} }
   & \colhead{HD or MHD}
}
\startdata 
KF20  & $512^3$ & $\sim$7 & 1  & 20                  & 0.084 & 0.014 & 90 & 1.6$\times 10^4$ &  1024 & 1.0 & MHD \\
KF2.5\_20 & $512^3$ & $\sim$1 & 0  & 2.5 \& 20  & 0.041  &  0.0033  & 16 & 510 & -& - & HD
\enddata
%%% 0.043  , 0.0037 ,  16 , 530 if we use spectra of V_c and Sigma
%% Text for table notes should follow after the \enddata but before
%% the \end{deluxetable}. Make sure there is at least one \tablenotemark
%% in the table for each \tablenotetext.
%\tablecomments{Table \ref{tbl-1} is published in its entirety in the 
%electronic edition of the {\it Astrophysical Journal}.  A portion is 
%shown here for guidance regarding its form and content.}
%\tablenotetext{a}{Sample footnote for table~\ref{tbl-1} that was generated
%with the deluxetable environment}
\tablenotetext{a}{The Sonic Mach number.}
\tablenotetext{b}{The Alfv\'en speed of mean magnetic field.}
\tablenotetext{c}{Central driving wavenumber.}
\tablenotetext{d}{Standard deviation of centroid velocity of \textit{small-scale} fluctuations. For KF2.5\_20,
          we calculate $\sigma_{V_c}$ using Fourier velocity and density modes with $k\geq 10$ (see text for details).}
\tablenotetext{e}{Standard deviation of column density  of \textit{small-scale} fluctuations. For KF2.5\_20,
          we calculate $\sigma_\Sigma$ using Fourier density modes with $k\geq 10$ (see text for details).}
\tablenotetext{f}{Amplitude of large-scale column density. See Equation (\ref{eq:lsg}).}
\tablenotetext{g}{Amplitude of large-scale centroid velocity.  See Equation (\ref{eq:lsg}).}
\label{table_1}
\end{deluxetable}

\end{document}